# Roundtable Gossip Algorithm: A Novel Sparse Trust Mining Method for Large-scale Recommendation Systems

Mengdi Liu[1], Guangquan Xu[1,*]

[1] Tianjin Key Laboratory of Advanced Networking (TANK), School of Computer Science and Technology, Tianjin University, Tianjin, China, 300350
*Corresponding author: losin@tju.edu.cn

**Abstract.** Cold Start (CS) and sparse evaluation problems dramatically degrade recommendation performance in large-scale recommendation systems such as Taobao and eBay. We name this degradation as the sparse trust problem, which will cause the decrease of the recommendation accuracy rate. To address this problem we propose a novel sparse trust mining method, which is based on the Roundtable Gossip Algorithm (RGA). First, we define the relevant representation of sparse trust, which provides a research idea to solve the problem of sparse evidence in the large-scale recommendation system. Based on which the RGA is proposed for mining latent sparse trust relationships between entities in large-scale recommendation systems. Second, we propose an efficient and simple anti-sparsification method, which overcomes the disadvantages of random trust relationship propagation and Grade Inflation caused by different users have different standard for item rating. Finally, the experimental results show that our method can effectively mine new trust relationships and mitigate the sparse trust problem.

## 1 Introduction

With the increasing availability of online business interaction and the rapid development of the Internet, we face an enormous amount of digital information. The Big Data Age has already arrived and has profoundly affected people's daily lives. The rapid growth of commodity types also meets the different needs of users. However, the Taobao index [1] shows that each commodity has just a few buyers and fewer effective comments. Searching for useful information from an enormous amount of data is similar to looking for a needle in a haystack. Large-scale recommendation systems are sought after by electronic commerce websites, identity authentication & security systems, search engines and other applications with big data challenges [2] [3] [36] [37] [38] [39] [40] [41] [42]. However, large-scale recommendation systems face sparse trust problem where there is significant lack of trust evidence.

The recommendation system is the typical application scenario of trust theory. The sparse trust problem is caused by CS problem and sparse trust evaluation. The CS problem which makes it hard to build trust relationship consists of user CS and item CS. And sparse trust evaluation is the sparsity of the user's original trust relationship. Therefore, substantial research efforts are focused on the sparsity of user evaluation data [4] and the CS problem. In the context of big data, the amount of available social information is far beyond the range of what individuals or systems can afford or handle, and use effectively, which is called information overload. The recommendation system solves this problem by filtering out noise to find the information that the user desires. Moreover, it can predict whether the user wants the information supplied by the recommendation system. Some existing trust models, such as the Personalized trust model [5], VoteTrust model [6], ActiveTrust model [7], swift trust model [8], and STAR: Semiring Trust Inference for Trust-Aware Social Recommenders [9], take advantage of overall ratings to assess the sellers' performance and ignore some latent information in textual reviews [10]. The Textual Reputation Model [11] improves upon the traditional model by calculating a comprehensive reputation score of the seller based on users' reviews [12] [13]. However, the recommendations that are made by these recommendation systems for inactive or new users are inaccurate, and sometimes no recommendation can be made. This is because the user evaluates or interacts with very few (or no) items [14].

Taobao is the most popular online shopping platform in China and has nearly 500 million registered users, more than 60 million visitors every day, and more than 800 million online products [15]. Therefore, the User-Item matrix size of Taobao's recommendation system is 60 million*800 million. It is very difficult and ineffective to use Spark (Apache Spark is a fast and universal computing engine designed for large-scale data processing) to calculate such a huge matrix. The lack of trust evidence due to CS and sparse trust evaluation further aggravates the difficulty in computing the matrix. Thus, the sparse evaluation and CS problems of recommendation systems are ultimately challenging due to the sparsity of data or information [16].

In this paper, a method of sparse trust mining is proposed to implement the anti-sparsification to improve the accuracy of recommendation system. The main contributions of this paper are as follows:

- defining the concept of sparse trust;
- providing a unified formal description of sparse trust;
- proposing a novel RGA for mitigating the sparse trust problem.

The sparse trust theory provides a research idea to solve the problem of sparse evidence in the large-scale recommendation system. Accordingly, research on sparse trust representation, evaluation, reasoning and prediction will greatly promote the development and evolution of trust theory and help improve the accuracy of recommendation system. Experiments show that the proposed method can extract latent trust relationships more efficiently and mitigate sparse trust problems at lower cost.

The remainder of the paper is organized as follows: Section 2 introduces the related work on the sparsity problem in the recommendation system. Section 3 describes the preliminaries regarding the relevant representation of sparse trust. Section 4 presents the RGA for the sparse trust mining method. Section 5 describes the evaluation metrics and provides performance analysis and comparative experiment studies. Section 6 presents the conclusions and future work.

## 2 Related Work

Typically, scholars use the trust relationship method to overcome the sparsity to obtain the consensus of most people. Guo et al. [17] incorporated the trust relationship into the SVD++ model, but this work relied only on explicit and implicit influences of social trust relationships. Real-world users are often reluctant to disclose information due to privacy concerns [18]. Zhong et al. [19] proposed a computational dynamic trust model for user authorization to infer the latent trust relationship. Yao et al. [20] considered the influence of the user’s dual identity as a trustor and a trusted person on trust perception recommendation to obtain latent association rules.

Some trust propagation methods have also been proposed to solve the sparse trust problem. Konstas [21] used a Random Walk with Restart to model the friendship and social annotations of a music tracking recommendation system. Chen [22] recommended communities through potential Latent Dirichlet Allocation (LDA) and association rule mining techniques. In [23], conceptual typology and trust-building antecedents were proposed in cloud computing.

Some of the existing trust models that are listed above have different advantages. However, none of the existing works fully solve the problem of data sparsity due to CS and sparse trust evaluation. In particular, many current proposals are unable to achieve anti-sparsification over highly sparse datasets and as a result, this problem becomes all the more prominent and acute in the big data environment. Moreover, there is no guarantee of prediction accuracy in such case. In this paper, the experiment indicates that, compared with current trust methods, RGA's trust prediction of anti-sparsification is more accurate in all users and CS environment, especially for highly sparse dataset. This is a reliable anti-sparsification method, and according to experiments, this algorithm has a certain degree of advantage in terms of stability.

## 3 Preliminaries

To better investigate the sparse trust problem, it is necessary to define the relevant representation of sparse trust. Generally, the sparsity of ratings for CS items is higher than 85% [24], where sparsity is the proportion of unknown relationships.

**Definition 1 (trust)**: Trust is an emotional tendency which the subject believes that the object is responsible and honest, and when the subject adopts the behavior of the object, it will bring him positive feedback. It is often described in the form of probability, and it is highly time-dependent and space-dependent.

**Definition 2 (sparse trust)**: In the era of big data, the probability of direct contact between the two entities is getting smaller. Sparse trust refers to the divisible trust relationship that is masked by data sparsity. A relationship that is not described by direct or expressive evidence is called “true” sparse trust. In addition, a relationship that includes ambiguous information or data noise masking evidence is known as "false" sparse trust.

The degree of trust is reflected by the emotional intensity and described by a probability [25]. Furthermore, the emotional intensity is proportional to the probability description. We denote the trust value by $P_{AB}$, which is between 0 and 1 and indicates the degree of trust that entity A has in entity B. A formal description of sparse trust is ternary form. For example: [beer, diapers, 0.73]. There are many more zero elements of the sparse trust matrix than non-zero elements, and ternary vectors are used to store sparse trust relationships, which can save storage space.

## 4 Round-table Gossip Algorithm(RGA)

In this section, we describe RGA. In roundtable gossip, the final trust value of the two unassociated entities is given by the sum of multiple mining paths, where each path computes intermediate entities of the transferred trust relationship by iteration. All related entities are from the same virtual community and interrelated. In Section 4.1, we describe the underlying works for the RGA. In Section 4.2, we describe the RGA algorithm for mining the trust relationships

### 4.1 Round-table Gossip

**Normalizing Sparse Trust**. The roundtable algorithm [26] was used to determine the attack attributes for the attack decision process of a combat game. It abstracts possible event state sets into a round table, which is the origin of the round table algorithm. The trust value of each entity on the round table does not attenuate, which truly reflects the subject's emotional tendencies toward the object. Inspired by the roundtable algorithm, we improved the algorithm to apply it to our sparse trust.

Entities used for trust mining have different trust degree in the same network community. Furthermore, different users have different criteria for item rating. For example, some people like rating high scores or low scores and some people only rate in a small range, which is called the Grade Inflation in data mining area [27]. In order to place all entities on a round table eliminate the impact of Grade Inflating, it is necessary to normalize them. We utilize the improved softmax function [28] $G_{ij}$ to normalize sparse trust：

$$G_{ij} = \begin{cases} \dfrac{e^{p_{ij}}}{\sum_{j \in I} e^{p_{ij}}}, if\ e^{p_{ij}} \neq 0 \\ 0, otherwise \end{cases} \tag{1}$$

where variable *I* stands for the network community. This function ensures that the sum of the trust values after normalization will be 1. Notice that if $e^{P_{ij}} = 1$, this trust value is 0, which indicates distrust. This is known as the sparseness of trust. In our work, we define the value of $G_{ij}$ in this case as zero, which is used to mine the trust value.

The normalized sparse trust values do not distinguish between an object with which subject did not interact at all or an object with which subject has had a reasonable degree of interaction. Moreover, if $G_{ij} = G_{ik}$, we know that entity *i* to entity *j* and entity *k* has the same sparse trust value, but we do not know if both of them are high-trust, or if both of them are low-trust. Maybe these two trust values come from different network communities, and different communities have different evaluation criteria. That is, these normalized sparse trust values are relative, but there is no absolute interpretation. After the normalization of the trust value, the relativity of it can still be subtly reflected. This manner of normalizing sparse trust values has been chosen because it allows us to perform computations without re-normalizing the sparse trust values at each iteration (which is prohibitively costly in the trust transitive process) as shown below. This also eliminates the impact of Grade Inflation and leads to an elegant algorithmic model. This calculation method is shown as Algorithm 1 lines 2-17.

**Transitiveness.** RGA is based on a trust transitive mechanism and aims at finding target entities' acquaintances. It makes sense to weight their opinions according to the trust that the subject places in them:

$$t_{ik} = \sum_{j \in I} g_{ij} g_{jk} \tag{2}$$

where $t_{ik}$ represents the trust that entity *i* places in entity *k*, which is determined by asking their acquaintances (have relation both with entity *i* and entity *k*, such as entity *j* in formula(2)). Note that there exists a straightforward probabilistic interpretation of this method of gossip transitive, which is similar to the Random Surfer model of user behavior based on browsing web pages [29].

The intermediate entities transfer the trust value for two entities that do not have a trust relationship. In our algorithm, each entity variable contains two quantities of information: an adjacency list of nodes that have a trust relationship with the entity node, which is denoted as $Trustlist(i)$, and a set of their trust values, which is denoted as $Data_set(g_{ij})$. We use depth-first search to find intermediate entities. The searched entities are marked and stored in the stack. This calculation method is shown in Algorithm 1 lines 18-21.

**Latent Trust Mining.** Generally, social relationships are divided into single-modal and multi-modal. Many scholars solve the sparsity problem of trust by referring to other sources of information (such as an inactive user's relationships with his or her friends). In our work, we consider the issue of latent trust relationship mining in multiple network communities (refers to the online communication space including BBS/forum, post bar, bulletin board, personal knowledge release, group discussion, personal space, wireless value-added service and so on). Mining latent trust relationships for entities from the same community is known as homogenous association rule mining, and mining of entities in different communities is referred to as heterogeneous association rule mining. In practice, physical entity users often cannot interact in all communities. By mining latent information about a community, it is possible to draw on and aggregate their trust in other communities, which is significant for anti-sparsification of a sparse trust matrix.

Next, we will discuss how to aggregate the trust relationships from different communities. The first thing that needs to be clarified is the way in which the relationship of trust is expressed. For each path from a complete entity *i* to entity *k*, the sparse trust value is calculated according to Formula (2) and stored in the corresponding matrix $T_{ik}^{(r)}$, where *r* represents the number of intermediate entities. The calculation method of latent trust mining is shown as Algorithm 1 lines 22-26.

### 4.2 Overview of Roundtable Gossip Algorithm

Here we describe the anti-sparsification method to compute the sparse trust values based on roundtable gossip algorithm. The homogenous entity association method between subjects and the heterogeneous entity association method between subject and object are used to solve the problem of multi-social relation data sparsity. In some cases, an entity may have an inter community association with another entity that resides in a different community. The algorithm aggregates multiple trust paths and each path involves gossip with multiple entities. An overview of RGA is shown in Algorithm 1. We initialize self-confidence value of entities as 1 (line 3) before normalizing the sparse trust to mine more trust relationships.

**Algorithm 1:** Roundtable Gossip Algorithm

**Input:** Ternary Form P: $[i, j, p_{ij}]$;

**Definitions:** Sparse Matrix: $S_Matrix$ ; Number of nonzero elements: $Num_{nz}$ ; Number of Entities: *m*; Trust value of entity *i* in entity *k* when gossiping through *r* intermediate entities: $t_{ik}^{(r)}$ ;

**Output:** Trust Matrix $T$ .

1:**Initial:** $S_Matrix(i, j) \leftarrow P_{ij}$ ;

2:**for** $i \leftarrow 1$ to m **do**

3: $S_Matrix(i,i) \leftarrow 1$ ; //Updata Self Confidence

4: **for** $j \leftarrow 1$ to m **do**

5: **if** $e^{p_{ij}} \neq 1$ **then**

6: $p_{i.} \leftarrow e^{p_{ij}}$ ; //Sum of Trust Values for Each Row

7: **end if**

8: **end for**

9:**end for**

10:**for** $i \leftarrow 1$ to m **do**

11: **for** $j \leftarrow 1$ to m **do**

**if** $e^{p_{ij}} \neq 1$ **then**
  $g_{ij} \leftarrow e^{p_{ij}} / p_{i.}$ ; //Compute Normalized Sparse Trust
**else** $g_{ij} \leftarrow 0$ ;
**end if**
**end for**
**end for**
$t_{ik}^{(0)} = g_{ij}$ ;
**repeat**
  $t_{ik}^{(r+1)} \leftarrow g_{i1}t_{1k}^{(r)} + g_{i2}t_{2k}^{(r)} + \cdots + g_{ij}t_{jk}^{(r)}$ ; //Transitiveness
**Until** $\left| Num_{nz}^{t^{(r+1)}} - Num_{nz}^{t^{(r)}} \right| < \varepsilon$ ;
**for** $j \leftarrow 1$ to $r-1$ **do**
  $t_{ik} \leftarrow t_{ik}^{(r)}$ ; //Aggregate the Mining Trust Relationship
**end for**
$T_{ik} \leftarrow t_{ik} / r$ ;
**return** $T_{ik}$ ;

# 5 Experiments

In this section, we will assess the performance of our algorithm in mining trust relationships, which target large-scale recommendation system. We conduct extensive experiments with synthetic and real-world datasets. The anti-sparsification ability of the RGA is achieved by calculating trust matrices of various sparsity degrees.

## 5.1 Dataset

The experimental datasets are mainly divided into two parts: The first part consists of real data from CiaoDVDs [31] and is used to evaluate our algorithm vertically in terms of updateability, validity and stability (Sections 5.3). The second part consists of two representative datasets, which are taken from popular social networking websites, including Douban (www.douban.com) [32] and Epinions (www.epinions.com) [30], and is used to sufficiently validate the performance of our proposed methods. The statistics of the three datasets are presented in Table 1.

The acquisition of a matrix with a different sparsity degree is very important for simulation purposes. We need to use a random function to determine whether all nonzero trust values are valid to better simulate trust matrices with different sparsity degrees (valid trust values are unchanged and invalid trust values are set to zero). In other words, the sparsity degree of the original dataset that we used is uniquely identified. We consider simulating big data matrices of different sparse degree, where a big data matrix is used to infer the trust relationship.

**Table 1.** Datasets Statistics from popular social networking websites

| Statistics | CiaoDVDs | Douban | Epinions |
|---|---|---|---|
| Num. Users | 7,375 | 129,490 | 49,289 |
| Num. Items | 99,746 | 58,541 | 139,738 |
| Num. Ratings | 278,483 | 16,830,939 | 664,823 |
| (Sparsity degree) | 0.0379% | 0.2220% | 0.0097% |
| Friends/User | 15.16 | 13.22 | 9.88 |

The number of nonzero elements in the initial dataset is relative (not absolute; the error range of sparsity degree is 0.0005%). In the processing of data, eight random function values for eight sparsity degrees were obtained through repeated experimental comparisons. The sparsity degree with corresponding random function values are shown in Fig. 1.

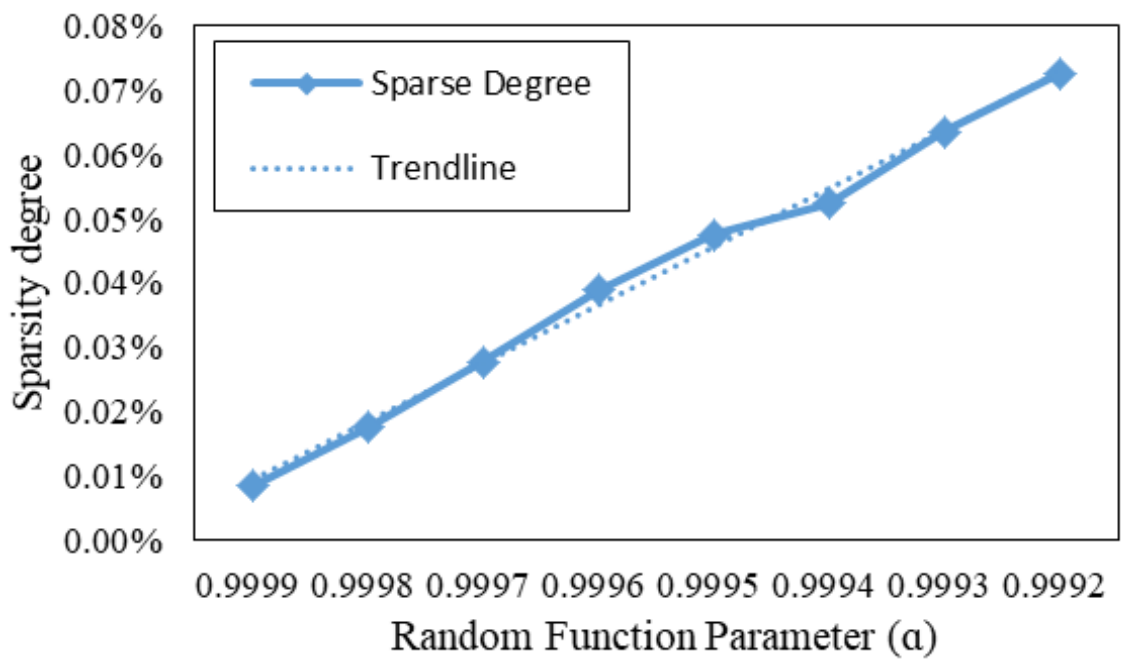

**Fig. 1.** Graph depicting of Sparsity Degree and Random Function. In the processing of data, eight random function values for eight sparsity degrees were obtained through repeated experimental comparisons.

### 5.2 Evaluation Metrics

In our experiment, we use two metrics, namely, the Mean Absolute Error (MAE) and the Root Mean Square Error (RMSE) [33], to measure the prediction quality of our proposed approach in comparison with other collaborative filtering and trust-aware recommendation methods.

The metric MAE is defined as:

$$MAE = \frac{\sum_{ij} \left| t_{ij} - \hat{t}_{ij} \right|}{N} \tag{3}$$

The metric RMSE is defined as:

$$RMSE = \sqrt{\frac{\sum_{ij} \left( t_{ij} - \hat{t}_{ij} \right)^2}{N}} \tag{4}$$

where $t_{ij}$ denotes the original trust value that entity $i$ gave to entity $j$, $\hat{t}_{ij}$ denotes the predict trust value that entity $i$ gave to entity $j$, as predicted by a method, and $N$ denotes the number of tested trust values.

### 5.3 Performance Analysis

We experimented with datasets of different sparsity degrees and compared the updateability, validity and stability of the algorithm in four different cases. The dataset describes the subject and object in ternary. A sparse matrix has a huge number of zero elements in a big data environment, so the use of ternary form $\left[i, j, p_{ij}\right]$ to store sparse matrices reduces spatial overhead and results in better trust delivery. The processing performed by the algorithm on matrices varies with different initial sparsity degree values. The processing ability of the algorithm is evaluated in terms of the following:

Updateability: The diagonal of a sparse trust matrix indicates the self-confidence of the entity that interacts in the recommendation system. In our scheme, we assume the entity has a self-confidence relationship and the value is equal 1, that is, the diagonal value of the sparse trust matrix is 1. Obviously, this initialization of the self confidence is helpful for anti-sparsification. Furthermore, even though the nonzero elements of the matrix reveal the relationship of the interacting entities, the role of the zero elements cannot be ignored. Our algorithm can establish a new relation, update the existing trust relationship, and provide more information for further association rule mining.

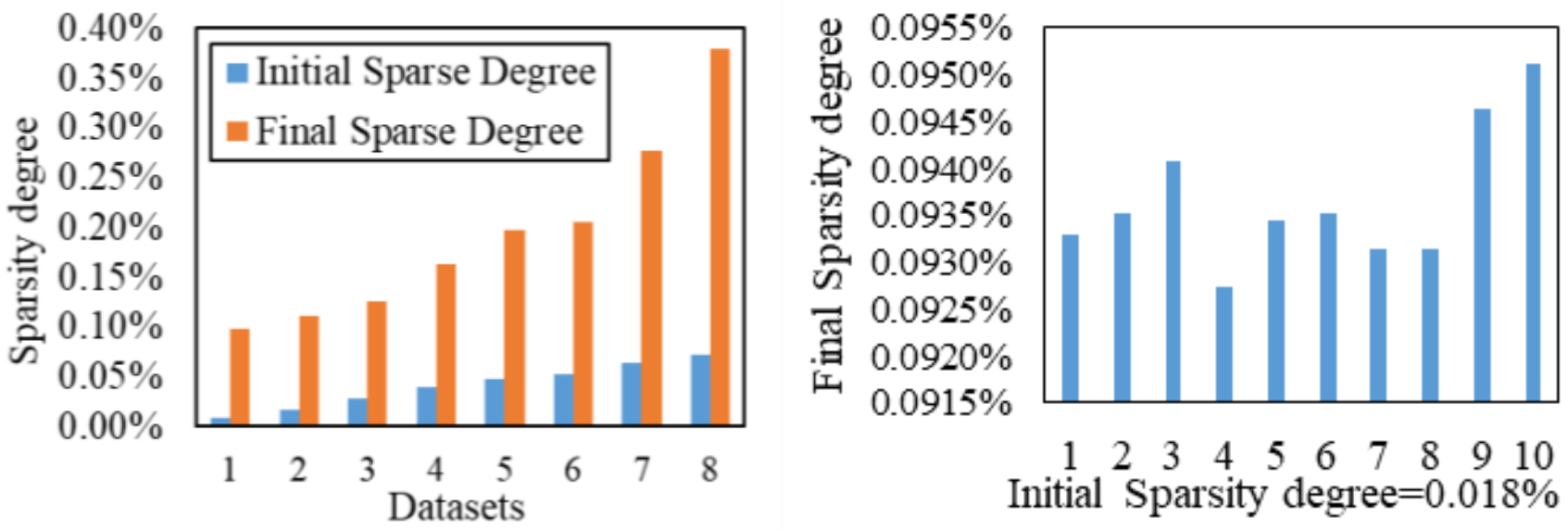

(a) (b)

**Fig. 2.** (a)Column chart of datasets of 8 different sparsity degrees, compared after anti-sparsification operations. (b)Column chart of 10 datasets' sparsity degrees after anti-sparsification operations, which have the same initial sparsity degree of 0.018%.

Validity: In this paper, experiments on 8 kinds of sparse matrices with different sparsity degrees are conducted to investigate the effectiveness of the proposed algorithm. The sparsity degrees (the proportion of nonzero elements to total elements) of the 8 matrices are as follows: 0.009%, 0.018%, 0.028%, 0.039%, 0.048%, 0.052%, 0.064%, and 0.072%. Obviously, these are sparse matrices [24]. Based on the probability description $t_{ik} = \sum_{j \in I} g_{ij} g_{jk}$ of the trust transitive mechanism, the intermediate entities transfer the trust value for two entities that do not have a trust relationship. In addition, the sum of multiple mining paths is expressed as the final trust, where each path computes intermediate entities of the transferred trust relationship by iteration. Fig. 2(a) gives a histogram that illustrates the anti-sparsification situation and each dataset corresponds to a sparse trust matrix with a different sparsity degree. The experimental data in Fig. 2(a) are the average values over many experiments, because each experimental result is influenced by many factors and single experiments are unreproducible (the error range of the sparsity degree is 0.0005%). Obviously, the algorithm can achieve anti-sparsification. In particular, the more is the evidence provided initially the greater the effect of the anti-sparsity. However, the sparse matrix itself can provide little evidence, which is a challenge for the experimental algorithms. Another challenge is the distribution of nonzero elements. We performed many experiments on trust matrices with the same sparsity degree and then selected ten representative experimental data with sparsity degree of 0.018%. As shown in Fig. 2(b), each column represents the residual $\left|Sparse_v^{'} - Sparse_v\right|$ of the sparsity degree. The results of the ten sets of experimental data are not the same because the different nonzero element locations lead to different anti-sparsification results. This is because the nonzero element locations determine the number of intermediate entities and the number of mining paths. However, for any initial sparse matrix, the RGA can maximize the anti-sparsity.

Stability: Algorithm validity is affected by the locations of sparse nodes, but an overall impact is minimal or negligible. The anti-sparsification algorithm is affected by two major challenges: One is the sparsity of trust matrix evidence, and the other is the different anti-sparsity effects of different sparse node locations with same sparsity degree. The latter will affect algorithm validity (as shown in Fig. 3). For example, a trust relationship is communicated by intermediate entities and the locations of nonzero elements determine the route of trust transmission, which directly affects sparse trust value computation by the RGA. By aiming to target this problem, we design a corresponding contrast experiment. We select two datasets in our contrast experiment, each of which contains 28 kinds of trust matrices with different sparsity degrees. The experimental results are shown in Fig. 3. The two curves represent two sets of comparative experiments, and each group consists of 28 sparse residuals for data with different sparsity degrees. The result shows that the sparse node locations affect the algorithm validity, but have little influence on the overall trend, so the algorithm is stable.

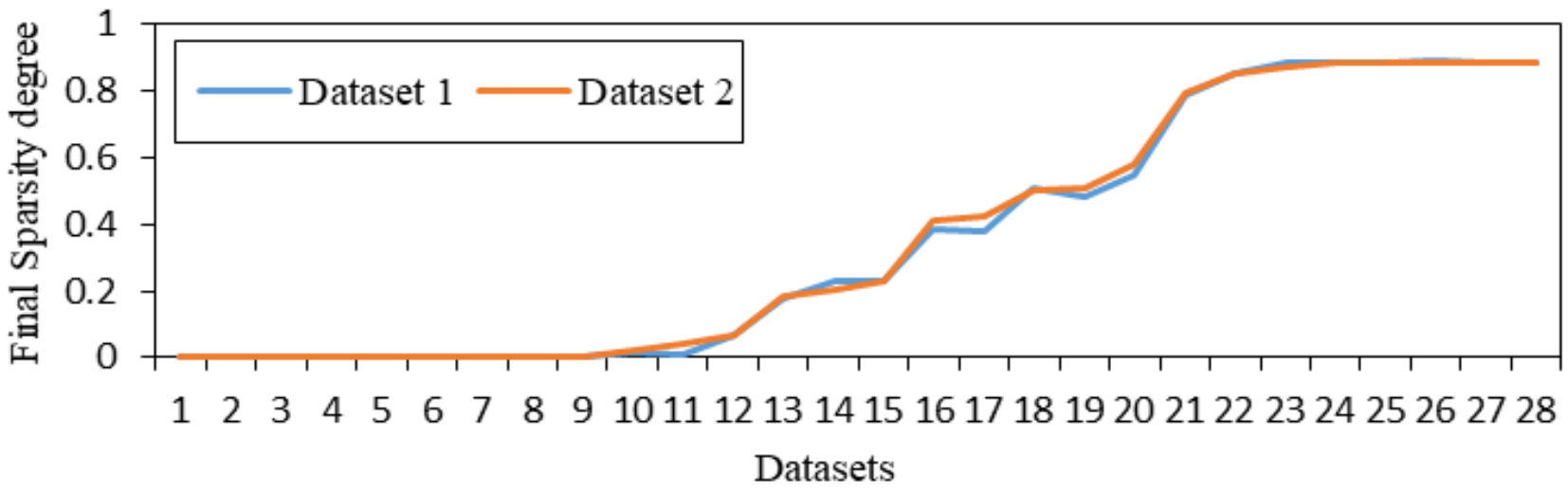

**Fig. 3.** Residuals of Anti-sparsity

### 5.4 Comparative Experimental Studies

In this section, to evaluate the performance improvement of our RGA approach, we compare our method with the following approaches:

**Comparison Methods.** To comparatively evaluate the performances of our proposed methods, we select three representative trust propagation methods as competitors: Mole Trust (MT), Propagation of trust (PT), and Tidal Trust (TT).

MT is able to propagate trust over the trust network and is capable of estimating a trust weight that can be used in place of the similarity weight [30]. PT develops a framework of trust propagation schemes [34]. TT presents two sets of algorithms for calculating trust inferences: one for networks with binary trust ratings, and one for continuous ratings [35]. All these methods use the trust transfer mechanism to predict trust relationship. The anti-sparsification accuracy of these method and RGA are verified by comparing experiments in all users and CS environment. To focus on verification and fair comparison, for all methods, we set the optimal parameters for each method according to their corresponding references or based on our experiments: Mole Trust: *mpd=Num.iterations*; Propagation of trust: $\alpha$ = (0.4, 0.4, 0.1, 0.1); Tidal Trust: *max*=0.008.

**Compared-Validation.** We employ the method of compared-validation for predicting and testing. We randomly divide the trust data into two equal parts: each time, we use one part as the predict set to predict sparse trust value (50 percent of the sparse trust data) and another part as the test set to compute MAE and RMSE (the remaining 50 percent of the sparse data). In addition, we conduct each experiment five times and take the mean as the final result for each experiment, as discussed below.

### 5.5 Results and Analysis

The anti-sparsification accuracy of the proposed RGA algorithm is verified by comparing the experimental results on the two datasets with those of the competing methods. Here, three algorithms are considered: Mole Trust (MT), Propagation of trust (PT), Tidal Trust (TT). MAE and RMSE, which are two benchmark error evaluation metrics, are used here. Table 2 and Table 3 respectively show the results of MAE and RMSE on testing of all users and on testing of CS, which were computed based on the user's predictions. In addition, Fig. 4 and Fig. 5 show the performance comparison histogram of the experiments.

**Table 2.** Experimental Results on Testing of All Users

| Datasets | Measure | MT | PT | TT | RGA |
|---|---|---|---|---|---|
| Douban | MAE | 0.9309 | 0.6525 | 0.8703 | 0.1267 |
| | (Improve) | 86.390% | 80.582% | 85.442% | - |
| | RMSE | 0.9582 | 0.5139 | 1.6258 | 0.0922 |
| | (Improve) | 90.378% | 82.059% | 94.329% | - |
| Epinions | MAE | 2.1507 | 2.6023 | 0.0700 | 0.0474 |
| | (Improve) | 97.796% | 98.179% | 32.286% | - |
| | RMSE | 16.9959 | 17.7774 | 0.1286 | 0.0480 |
| | (Improve) | 99.718% | 99.730% | 62.675% | - |

**Validation on All Users.** RGA outperforms other approaches in terms of both MAE and RMSE on two datasets. PT method achieves the second-best performance on the two datasets in terms of MAE, except dataset Epinions. Because the trust data of Epinions is extremely sparse (sparsity degree of 0.0097%), the traditional proposed methods, namely, TT, performs much worse than the proposed methods, namely, PT and MT. However, when the trust data are relatively dense, such as in the Douban (density of 0.2220 percent), PT shows a comparable, and sometimes better, performance. Finally, for Epinions, which contains directed trust networks, TT is more accurate than PT or MT. However, for Douban, which contain undirected friend networks, the performance of MT is similar to that of TT. Hence, the recommendation quality improvement that results from their combination is limited. Based on the above points, PT performs optimally on these series.

Furthermore, the improvements against respective competitors on testing of all users, which are given in the Table 2, show that our methods can significantly improve the quality of recommendations, especially for Epinions, which is a highly sparse dataset as described above in Section 5.1. Experimental result on testing all users proves that on the testing of all users RGA can be implemented anti-sparsification. This is a reliable anti-sparsification method, and according to several experiments, this algorithm has a certain degree of advantage in terms of stability.

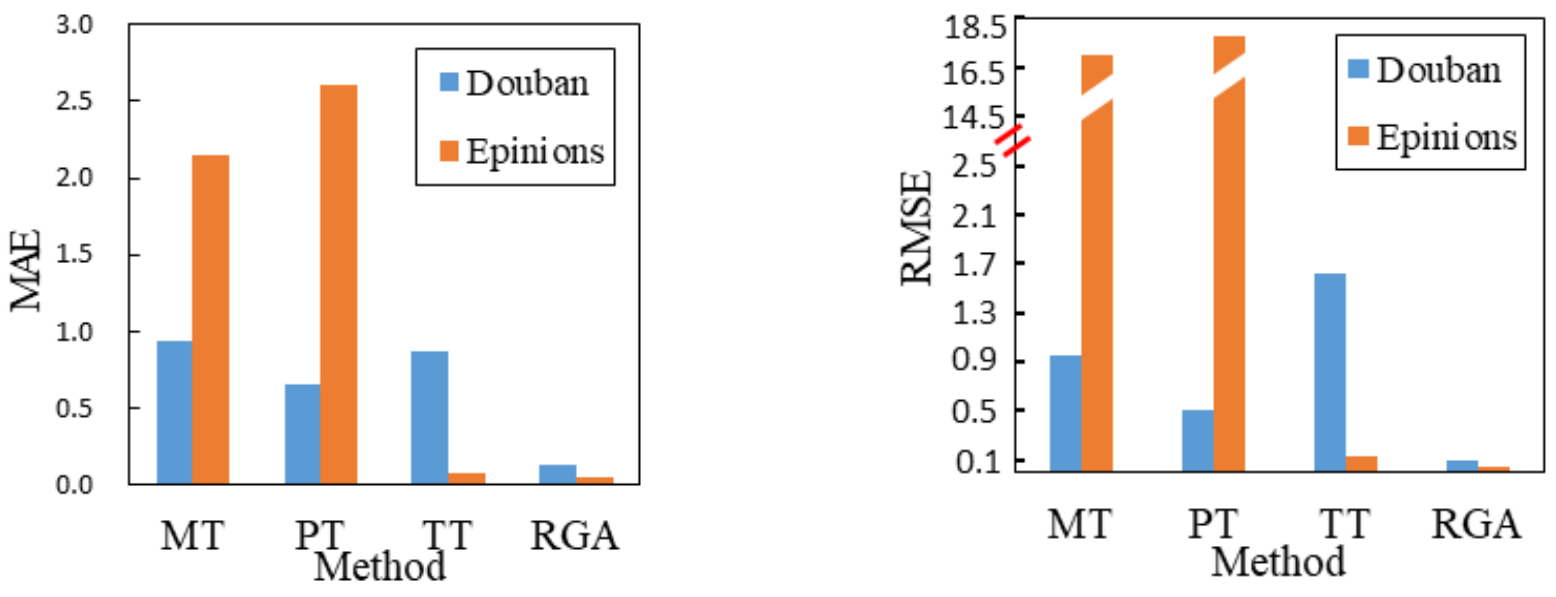

(a)MAE Comparison on Testing of All Users (b)RMSE Comparison on Testing of All Users

**Fig. 4.** Performance Comparison on Testing of All Users. Instructions: In order to improve the display effect of the data (the numerical gap is large), the truncation diagram is adopted in (b).

**Validation on CS Users.** As mentioned in Introduction, The CS problem becomes more severe and frequent in the large-scale recommendation system environment, which results in the sparse trust problem.

**Table 3.** Experimental Results on Testing CS User

| Datasets | Measure | MT | PT | TT | RGA |
|---|---|---|---|---|---|
| Douban | MAE | 0.9412 | 0.6742 | 0.9016 | 0.1495 |
| | (Improve) | 84.12% | 77.83% | 83.42% | - |
| | RMSE | 0.9603 | 0.6539 | 1.6473 | 0.1103 |
| | (Improve) | 88.51% | 83.13% | 93.30% | - |
| Epinions | MAE | 2.1705 | 2.7183 | 0.0918 | 0.0714 |
| | (Improve) | 96.71% | 97.37% | 22.22% | - |
| | RMSE | 17.1202 | 17.97474 | 0.1416 | 0.0606 |
| | (Improve) | 99.65% | 99.66% | 57.20% | - |

For this, we also evaluated the accurate performance of the RGA's anti-sparsification of trust relationship in CS environment. Generally, we define the users who have fewer than five trust relationships as CS user. Compared-validation is still used in the test but we only care about the accuracy of predictions for CS users (with five or fewer trust relationships) at this moment. Table 3 shows that RGA still have the best performance on dataset Douban and Epinions, especially for highly sparse dataset Epinions, which proves that RGA can be implemented accuracy anti-sparsification on the testing of CS.

The experiment indicates that, compared with current trust methods, RGA's trust prediction of anti-sparsification is more accurate in all users and CS environment.

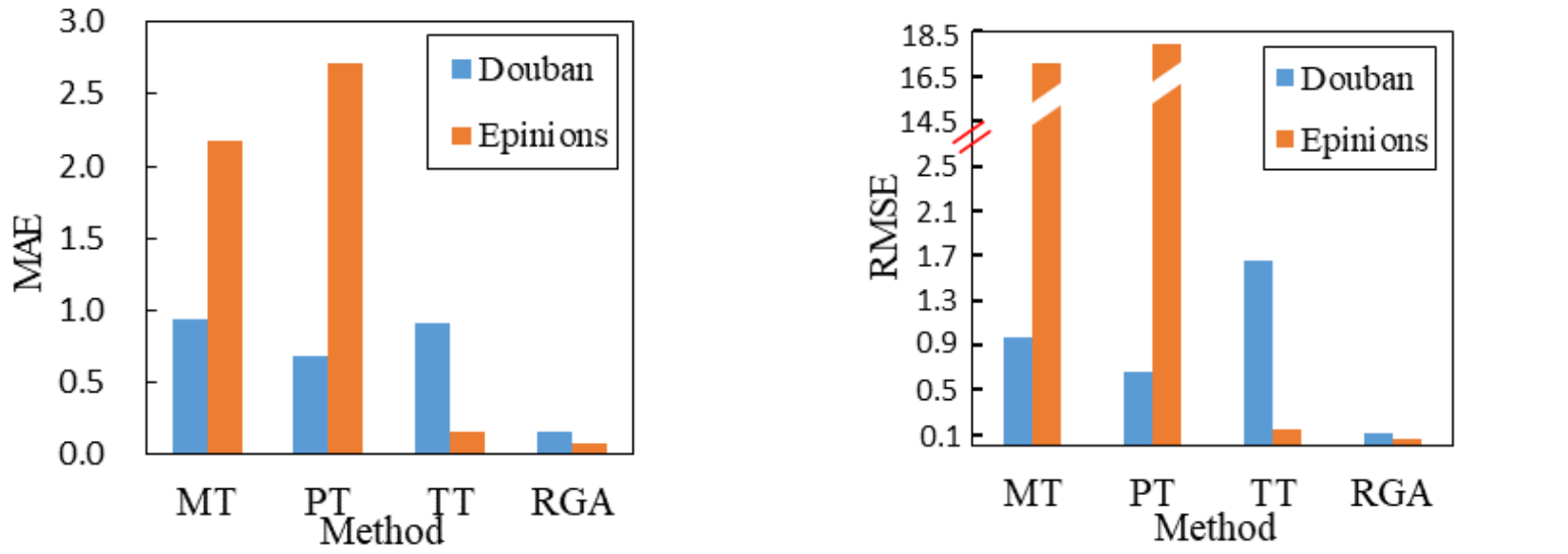

(a)MAE Comparison on Testing of CS (b)RMSE Comparison on Testing of CS

**Fig. 5.** Performance Comparison on Testing of CS. Instructions: In order to improve the display effect of the data (the numerical gap is large), the truncation diagram is adopted in (b).

## 6 Conclusions and Future Work

In this paper, we presented a novel method for mining and predicting sparse trust relationships in a large-scale recommendation system. Our proposed sparse trust mining method achieves anti-sparsification for sparse trust relationship, which is mainly based on RGA. Thus, by taking into account the trust transfer relationship, we also show how to carry out the computations in a big data environment. Sparsity degree measurement is used to analyze the anti-sparsification performance. In addition, this method updates the trust relationship in an effective and scalable way. However, the sparsity degree of the data that are processed in the experiments may exceed the requirement that it should be less than one millionth [21], and although the trust relationship accuracy has been preliminarily measured by MAE and RMSE, we can also mine the hidden semantics to improve the prediction accuracy and effectiveness. These problems need to be considered in future work.

## Acknowledgement

This work has been partially sponsored by the National Science Foundation of China (No. 61572355, U1736115), the Tianjin Research Program of Application Foundation and Advanced Technology (No. 15JCYBJC15700), and the Fundamental Research of Xinjiang Corps (No. 2016AC015).